\documentclass[%
 prl,
 amsmath,amssymb,
 reprint,%
floatfix,
superscriptaddress
]{revtex4-1}

\usepackage{graphicx}
\usepackage{dcolumn}
\usepackage{bm}

\usepackage[utf8]{inputenc}
\usepackage[T1]{fontenc}
\usepackage{mathptmx}
\usepackage{etoolbox}

\usepackage{xcolor,hyperref}
\hypersetup{
   colorlinks,
   linkcolor={blue!50!black},
   citecolor={blue!50!black},
   urlcolor={blue!80!black}
}
\makeatletter
\def\@email#1#2{%
 \endgroup
 \patchcmd{\titleblock@produce}
  {\frontmatter@RRAPformat}
  {\frontmatter@RRAPformat{\produce@RRAP{*#1\href{mailto:#2}{#2}}}\frontmatter@RRAPformat}
  {}{}
}%
\makeatother

\draft 

\begin{document}


\title{Anomalous softness in amorphous matter in the reversible plastic regime} 


\author{A.~Elgailani}
\affiliation{Department of Mechanical and Industrial Engineering,
Northeastern University, Boston, Massachusetts 02115, USA}

\author{D.~Vandembroucq}
\affiliation{PMMH, CNRS UMR 7636, ESPCI Paris, PSL University, Sorbonne  Université, Université Paris Cité, F-75005 Paris, France}

\author{C.E. ~Maloney}
\affiliation{Department of Mechanical and Industrial Engineering, Northeastern University, Boston, Massachusetts 02115, USA}


\date{\today}

\begin{abstract}
We study an integer automaton elasto-plastic model of an amorphous solid subject to cyclic shear of amplitude $\Gamma$.
We focus on the reversible plastic regime (RPR) at intermediate $\Gamma_0<\Gamma<\Gamma_y$, where, after a transient, the system settles into a periodic limit cycle with hysteretic, dissipative plastic events which precisely repeat after an integer number of cycles.
We study the plastic strain rate, $\frac{d\epsilon}{d\gamma}$, (where $\gamma$ is the applied strain and $\epsilon$ is the resulting plastic strain) during the terminal limit cycles and show that it consists of a creeping regime at low $\gamma$ with very low $\frac{d\epsilon}{d\gamma}$ followed by a sharp transition at a characteristic strain, $\gamma_*$, and stress, $\sigma_*$, to a flowing regime with higher $\frac{d\epsilon}{d\gamma}$.
We show that while increasing $\Gamma$ above $\Gamma_0$ results in lower terminal ground state energy, $U_{\text{min}}$, and a correspondingly narrower distribution of stresses, it, surprisingly, results in \emph{lower} $\gamma_*$, and $\sigma_*$.
The stress distribution, $P(\sigma)$, also becomes skewed for $\Gamma>\Gamma_0$.
That is, the systems in the RPR are anomalously soft and mechanically polarized.
We relate this to an emergent characteristic feature in the stress distribution, $P(\sigma)$, at a value, $\sigma_0$, which is independent of $\Gamma$ and show that $\sigma_0$ implies a relation between the $\Gamma$ dependence of $\sigma_*$, $\gamma_*$, and the amplitude of plastic strain, $\epsilon_p$.
We show that the onset of hysteresis is characterized by a power-law scaling, indicative of a second order transition with $\epsilon_p\propto (\Gamma-\Gamma_0)^{1.2\pm0.1}$.
We argue that $\sigma_0$ and, correspondingly, the onset of the RPR at $\Gamma=\Gamma_0$, is simply set by the so-called Eshelby-stress; the stress a local shear transformation zone will find itself under after a plastic transformation.
Furthermore, we show that cycling at $\Gamma_0$ results in a state which is maximally hardened.
Our results place important constraints on mean-field descriptions of cyclic yielding and should motivate experiments and particle-based simulations on cyclically sheared systems such as amorphous alloys, colloidal glasses, emulsions, pastes, granular matter, etc.
\end{abstract}

\pacs{}

\maketitle 

Most of the materials around us are strongly out of equilibrium.
Their properties, therefore, depend on how they came into being.
Examples range from polycrystalline alloys to thermoplastics to the soils we build our houses on and the wood and concrete we use to build them.
In particular, it is well known that allowing an amorphous alloy~\cite{Sun:2016we,Das:2019th,Ruta:2012wm,Ruta:2017wi,Evenson:2015ub,PhysRevB.89.174204,Fan:2014vf} or a glassy colloidal suspension~\cite{Bonn:1999vw,Abou:2001te,Rogers:2008wb} to age will generally result in a decrease in the free energy, an increase the shear modulus, and a decrease in the amount of plasticity the material would exhibit when subjected to loading.
That is, aging, or thermal annealing generically hardens these materials.
Mechanical preparation protocols may also be employed, and have been used since ancient times(~\cite{Huang:2018us}).
In fact, in athermal or non-Brownian systems like foams, emulsions, or granular matter, mechanical processing is the \emph{only} way to change the state of the material.
However, in mechanical preparation protocols, the situation is a bit more complex than in the thermal case and depends on the amplitude of the mechanical perturbation.
For small perturbations, one generally assists the thermal annealing and further lowers the free energy of the system, while for larger amplitudes, one will “rejuvenate” the glass and send it back to states with higher energy reminiscent of younger glasses with less aging~\cite{Lacks-PRL04,Warren:2008wm,Xue:2022wq,Kaloun:2005vn, McKenna:2003tk, Cipelletti:2005uj, Viasnoff:2002wa,Di-Dio:2022uu,Cloitre:2000wu,Ozon:2003uq,Viasnoff:2003vj,Cohen-Addad:2001vl,Narita:2004wb,Sollich:1997up,Coussot:2002vx}.
One would expect — in analogy with the thermal case — that the lower the energy of the mechanically annealed states, the harder they would be.
Here we show, surprisingly, this is not always true.

Perhaps the simplest mechanical preparation protocol is cyclic simple shear in the forward and reverse sense. Cyclic shearing experiments have been performed on foams~\cite{lundberg2008reversible}, emulsions~\cite{Knowlton-SM2014}, microgel pastes~\cite{Mohan:2013ud}, and colloidal glasses~\cite{Mukherji:2019wp,Hima-Nagamanasa:2014vz,PhysRevE.66.051402}. A particularly clear, precise, and informative set of experiments has been performed by Keim and Arratia, and co-workers on 2D rafts of polystyrene particles trapped at an oil-water interface~\cite{PhysRevResearch.2.012004,PhysRevLett.112.028302,C3SM51014J,Galloway:2020ug,Galloway:2022tk,Farhadi:2017uo}. In those experiments, it was shown~\cite{PhysRevLett.112.028302,C3SM51014J} that, depending on the shearing amplitude, $\Gamma$, one reaches one of three terminal steady state behaviors. At the lowest $\Gamma$, $\Gamma<\Gamma_0$, after a transient, the system responds completely elastically with no rearrangements or any kind of energy dissipating events at all. We refer to this as the elastic regime (ER). For $\Gamma_0<\Gamma<\Gamma_y$, rearrangements — along with their associated energy dissipation and hysteresis — occur, but the rearrangements precisely reverse themselves microscopically after one or more shear cycles. The system remains trapped within a finite set of microscopic configurations, and there is no long time diffusive behavior despite the hysteresis and energy dissipation during the shear cycles. Following Keim and Arratia, we refer to this as the reversible plastic regime (RPR). Finally, for $\Gamma>\Gamma_y$, the system never returns to a previous configuration even after an arbitrarily large number of shear cycles, and the long-time behavior is diffusive. We refer to this regime as the diffusive regime (DR).


The reversible plastic behavior has also been observed in particle-based simulations.
Early simulations from Lundberg {\it et. al.}~\cite{lundberg2008reversible} demonstrated the existence of reversible plastic events.
Later work by Fiocco et. al.~\cite{fiocco2013oscillatory} clearly demonstrated both the onset of hysteretic and diffusive behavior but did not differentiate between the two, and, in this sense, was blind to the possibility of reversible plasticity.
Priezjev showed evidence for a reversible plastic regime~\cite{priezjev2013heterogeneous,Priezjev:2016wq}, but it was Regev and collaborators who first showed that the reversible trajectories  took the form of complex limit cycles with period greater than one~\cite{regev2013onset}.
Some previous work did not distinguish between the onset of hysteresis and the onset of irreversible diffusion and thus failed to identify reversible plasticity~\cite{kawasakiBerthier2016,doi:10.1063/5.0085064,fiocco2013oscillatory,  Leishangthem:2017us, doi:10.1073/pnas.2100227118},
but most recent work has focused on characterizing the reversible plastic limit cycles and explaining their origin~\cite{Mungan-PRL19,Szulc:2022uc,Mukherji:2019wp,Keim:wi,Khirallah-PRL2021, Regev:2019ut, Regev-PRE20-energy-landscape,PhysRevE.99.052132,doi:10.1126/sciadv.abg7133,Keim:vj,PhysRevResearch.2.012004,Terzi:2020td,Keim:2019vm,Otsuki:2022tx,C9SM01488H}(see~\cite{https://doi.org/10.48550/arxiv.2211.03775} for a recent review).

In terms of the ER->RPR and RPR->DR  transitions, most previous work, including our own~\cite{Khirallah-PRL2021,doi:10.1063/5.0102669}, has focused on  RPR->DR~\cite{Regev:2019ut, https://doi.org/10.48550/arxiv.2211.03775,PhysRevE.103.062614,Regev:2015wt,regev2013onset, Mungan-PRL19,Regev-PRE20-energy-landscape,Priezjev:2016wq}. 
Here, we focus instead on the ER->RPR transition -- which we show below to have a mixed first-second order character.
Interestingly, we show that this transition is associated with an optimal hardness, and we explain the location of the transition, $\Gamma_0$, in terms of the so-called Eshelby stress in our model.
For amplitudes in the ER where the terminal limit cycles are trivial, larger amplitudes result in lower energy and harder terminal states as one would naively expect in analogy with thermal annealing.
In contrast, in the RPR where the terminal limit cycles are plastic and hysteretic, larger amplitudes result in softer terminal states \emph{despite} the lower energy.
Furthermore, the RPR states become increasingly mechanically polarized with skewed distributions of stress.

We employ a simple elasto-plastic integer automaton model (EPM) for an amorphous solid subjected to cyclic shear in the athermal, quasi static (AQS) limit.
EPMs are based on the notion of localized yielding events coupled with non-local stress redistribution after yield.
The model and the initialization procedure are described in detail in our previous work\cite{Khirallah-PRL2021} and the appendix, but the important feature which distinguishes it from other related EPMs~\cite{Tyukodi-PRL18,ISI:000491996300004,nicolas2018deformation,Liu:2022tb} is that we introduce no disorder or stochasticity into the model after an initial dynamical quench protocol which allows capturing the periodic limit cycles in the RPR.
In our EPM, we have an additive decomposition of the strain into an elastic and plastic contribution so that $\sigma=\mu (\gamma-\epsilon)$ where $\sigma$ is the total shear stress, $\gamma$ is the imposed strain, and $\epsilon$ is the plastic strain in the material.
$\mu$ is the elastic shear modulus, and we measure all stresses in this work in units of $\mu$, so we simply have: $\sigma=\gamma-\epsilon$.

In the following, we will show that if we characterize the system via the value of the ground state energy, the transition at $\Gamma_0$ is essentially undetectable.
However, if we characterize the system via the plastic strain amplitude in the terminal limit cycle, then we have something like a second order transition, where below $\Gamma_0$, there is precisely zero energy dissipated and no plastic strain, but above $\Gamma_0$, the dissipated energy and the plastic strain, $\epsilon_p$ increase with $\Gamma$ continuously from zero as a powerlaw with $\epsilon_p\propto (\Gamma-\Gamma_0)^{1.2\pm 0.1}$.
If we look at the shape of the $\epsilon$ vs $\gamma$ hysteresis curves, above $\Gamma_0$, the plastic strain rate, $\frac{d \epsilon}{d\gamma }$ turns on quite suddenly at a characteristic strain value, $\gamma_*$, and jumps from zero to a finite value of about $0.15$ regardless of the cycling amplitude; reminiscent of a first order transition.
Although the value of the plastic strain rate at the onset of plasticity is insensitive to $\Gamma$, the strain and stress at which plasticity begins, $\gamma_*$ and $\sigma_*$ both \emph{decrease} with increasing $\Gamma$ indicating that the systems which are cycled at higher $\Gamma$ are actually \emph{softer} than those cycled at lower $\Gamma$.
This behavior is striking and counter-intuitive, as one would expect the systems at higher $\Gamma$ to be \emph{harder} given their lower energy and narrower stress distribution…. presumably indicative of a deeper quench.
We show that the $\Gamma$ dependence of $\gamma_*$, $\sigma_*$, and $\epsilon_p$ can be understood by the emergence of a non-trivial $\Gamma$-\emph{independent} characteristic stress, $\sigma_0$ which is evident in the probability distribution of the stress, $P(\sigma)$, for all systems in the RPR.  

\begin{figure}[h!]
\includegraphics[width = 1.\columnwidth]{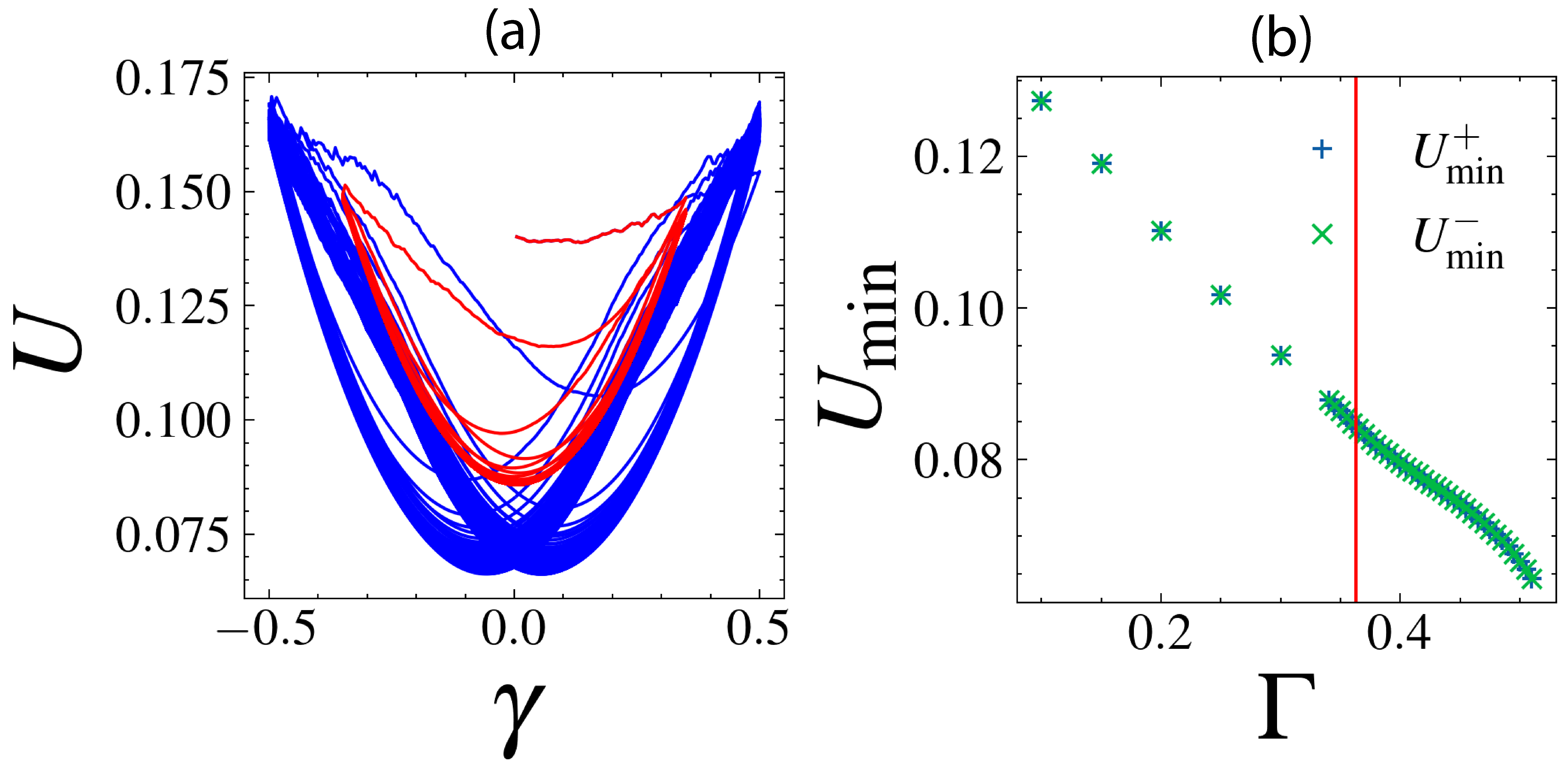}%
\caption{a) Energy U vs strain $\gamma$ for amplitudes $\Gamma=0.35$ in the ER (red) and $\Gamma=0.5$ in the RPR (blue). 
b) The minimum energy in the terminal limit cycle, $U_{\text{min}}^{\pm}$, in the forward and reverse direction vs $\Gamma$. 
The red vertical line is at $\Gamma_0=0.36320$ defined in the text.}
\label{fig:fig2}
\end{figure}
In figure~\ref{fig:fig2}a, we plot $U$ vs. $\gamma$, during cycling for two typical amplitudes, one at $\Gamma=0.35 < \Gamma_0$ in the ER just below the onset of reversible-hysteretic behavior, and one at $\Gamma=0.5 > \Gamma_0$ well into the RPR obvious hysteresis.
The ER system terminates in a trivial single-valued $U$ vs. $\gamma$ curve, while the RPR system terminates in a double-valued hysteretic curve 
\footnote{Note that while the terminal limit-cycles in the ensemble may have different periods, we perform an ensemble average, so the curves are precisely \emph{double}-valued.}.
We note that even though the final steady states in the ER are trivial and devoid of any plastic relaxation, the $U_\text{min}$, the energy of the stress-free configuration visited during a shear cycle, decreases with each successive cycle until the plasticity vanishes (see Fig.~\ref{fig:fig2}), so the terminal ground state energy is a non-trivial function of $\Gamma$, even in the ER.  
In figure~\ref{fig:fig2}b, we show the energy minimum, $U_\text{min}^{\pm}$, in the forward and reverse directions in the terminal limit cycles.
In principle, $U_\text{min}^{\pm}$ could be different from each other, and this is obviously so during the transient, but we find the difference negligible in the terminal limit cycles.
$U_\text{min}$ decreases montonically with increasing $\Gamma$ over the range of $\Gamma$ studied here.
There are no obvious features appearing at the onset of terminal hysteresis at $\Gamma_0\approx 0.365$.

\begin{figure}[h!t]
\includegraphics[width = 1.\columnwidth]{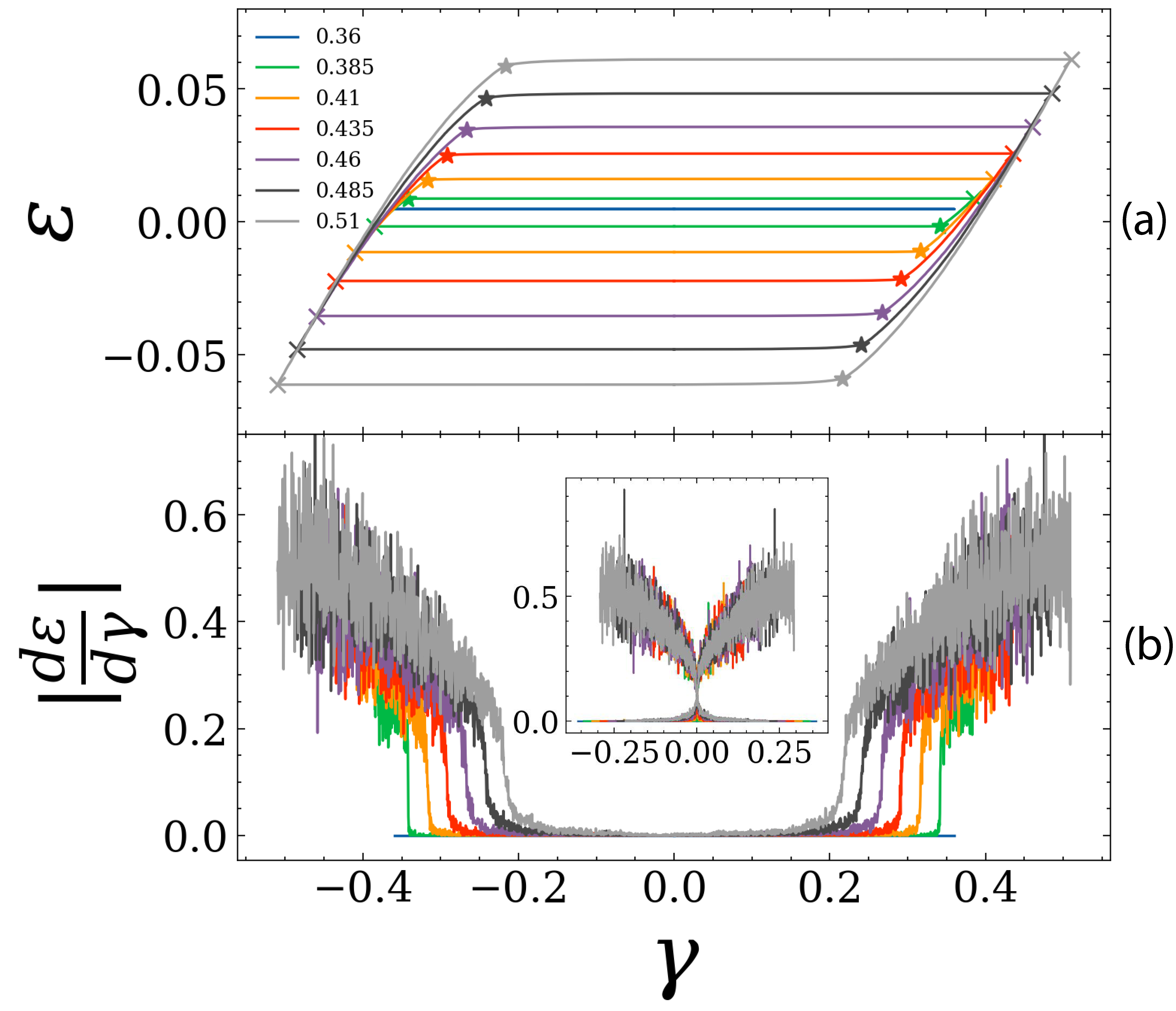}%
\caption{a) Plastic strain $\varepsilon$ (ensemble average) vs. strain $\gamma$ for different strain amplitudes (color online) in the terminal limit cycles. 
Crosses mark the turning points from forward to reverse driving, and stars mark, $\gamma_*$, the nominal onset of plasticity defined by $\left| \frac{d\epsilon}{d\gamma}\right|>0.1$.
b) Plastic strain rate, $\frac{d\epsilon}{d\gamma}$ vs $\gamma$.  The inset shows shifted data plotted as $\frac{d\epsilon}{d\gamma}$ vs $\gamma-\gamma_*$.
}
\label{fig:fig1}
\end{figure}
In figure~\ref{fig:fig1}, we show, in the terminal limit cycles, the ensemble average plastic strain, $\epsilon$, vs. applied strain, $\gamma$, and its derivative, the plastic strain rate, $\frac{d\epsilon}{d\gamma}$ for several different cycling amplitudes, $\Gamma$
\footnote{We count each strain-sweep in any given limit cycle independently in our ensemble of hysteresis loops; e.g. if a system has a period $5$ limit cycle, each of the $5$ strain-sweeps counts independently in the ensemble average.}.
The curve at the lowest amplitude of $\Gamma=0.36$ is in the elastic regime where all terminal limit cycles in the ensemble are trivial and devoid of plasticity.  
After some initial plasticity during the first several cycles, the plasticity terminates, so the terminal $\epsilon$ vs $\gamma$ plot is simply a flat line.
Nevertheless, due to the plastic activity during the initial transient cycles, there is a residual plastic strain in the positive sense, and the flat line sits at a positive $\epsilon$ value. 
This symmetry breaking is a result of the initial choice to first push the system in the forward direction to begin cycling.
At higher $\Gamma$, the terminal limit cycles are non-trivial, and there is a proper hysteresis loop with associated energy dissipation. 
Within a limit cycle of amplitude $\Gamma$, as the strain, $\gamma$, increases, starting from $-\Gamma$ toward $+\Gamma$, the plastic strain values, $\epsilon$, are initially perfectly flat indicating absence of any plastic activity.
However, as $\gamma$ increases, eventually significant plasticity initiates and the $\epsilon$ increases with $\gamma$.
Clearly, even the systems at large $\Gamma$ are still quite far from reaching steady state before the turning point at $\gamma=\Gamma$.

In figure~\ref{fig:fig1}b, we show the plastic strain rate, $\frac{d\epsilon}{d\gamma}$.
The $\frac{d\epsilon}{d\gamma}$ curves make it clear that, for any amplitude, the initiation of plastic activity is rather sharp, and furthermore, the plastic strain rate jumps to a value of approximately $0.20$ regardless of cycling amplitude.
It is also striking -- and it is the central result of this letter -- that the systems cycled at higher amplitude exhibit the onset to plasticity at \emph{lower} values of strain. 
To define the onset of plastic flow, we use the condition $\frac{d\epsilon}{d\gamma}>0.1$ and denote the value of $\gamma$ at which the transition occurs as $\gamma_*^{\pm}$ for the forward and reverse strain sweeps respectively.
In the inset, we plot $\frac{d\epsilon}{d\gamma}$ as a function of $\gamma-\gamma_*$ which shows that, to a good approximation, we can view $\frac{d\epsilon}{d\gamma}$ as an amplitude independent function of $\gamma-\gamma_*$.
It is also clear that, while all systems have a reasonably sharp jump from a small $\frac{d\epsilon}{d\gamma}$ to roughly $0.2$, the systems at smallest amplitude have the sharpest jump with significantly more rounding for the larger amplitude systems.
This regime of small but non-zero values of $\frac{d\epsilon}{d\gamma}$ for $\gamma<\gamma_*$ is reminiscent of below-threshold thermally assisted creep in complex fluids~\cite{Bouttes:2013th, Merabia:2016um, Castellanos:2019ur, Liu:2021un, Ferrero:2021ta, Popovic:2022wt} and thermal rounding in depinning phenomena~\cite{Caballero:2018vz,Middleton:1992vh, Bustingorry:2008te, Kolton:2020tu, Prevost:1999ws, Paruch:2013vc, Ferre:2013tl}, but here the creep arises from bona fide athermal yielding of weak sites with no thermal activation per se.


\begin{figure}[h!t]
\includegraphics[width = 1.\columnwidth]{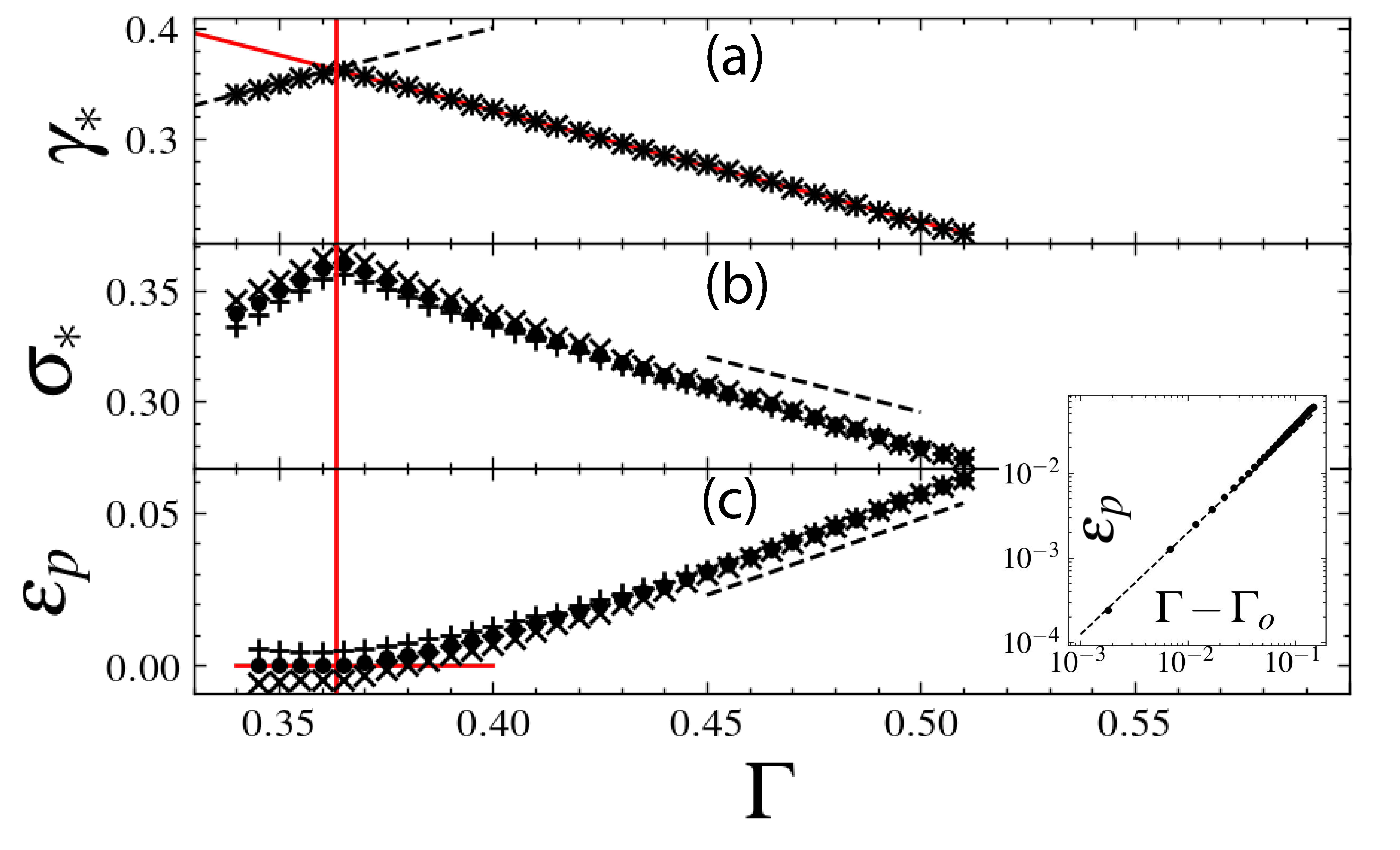}%
\caption{a) The (negative) onset strain, $\pm\gamma_*^{+(-)}$, in the forward (and reverse) direction.
The red line is $\gamma_*=\Gamma_0-(\Gamma-\Gamma_0)$ with $\Gamma_0=0.36320$, as determined by taking an average of $(\gamma_*+\Gamma)/2$.
b) The (negative) stress at onset, $\pm\sigma_*^{+(-)}$, in the forward (and reverse) direction.
c) The (negative) plateau strain, $\pm\epsilon_p^{+(-)}$, in the forward (and reverse) direction.
For all cases, the pluses are for the forward direction, the crosses symbols are for the reverse direction, and for b) and c), the averages are indicated by dots.
The dashed lines in b) and c) are guides to the eye with slope $\pm 1/2$ respectively.
The red vertical line is at $\Gamma_0=0.36320$. 
The red horizontal line is at $\epsilon_p=0$.
Inset: $\epsilon_p$ vs $\Gamma-\Gamma_0$ on a log-log scale with the dashed line indicating $\epsilon_p \propto (\Gamma-\Gamma_0)^{1.2}$. }
\label{fig:fig3}
\end{figure}
In figure~\ref{fig:fig3}a we show the yield point, $\gamma_*$ vs $\Gamma$.
If any of the systems in the ER, initially cycled at $\Gamma_1$, were suddenly subjected to cycling at higher amplitude, $\Gamma_2$, with $\Gamma_2>\Gamma_1$, those systems would exhibit a sudden burst of plasticity at $\gamma=\Gamma_1$.
Therefore, we would obtain $\gamma_*=\Gamma$ for systems in the ER.
At the onset of the RPR where $\Gamma=\Gamma_0$, $\gamma_*$ is essentially equal to $\Gamma$ indicating that there is essentially no plasticity until reversal.
Beyond $\Gamma_0$, we see $\gamma_*$ almost perfectly follow a straight line with slope of $-1$.  
That is: $\gamma_*=\Gamma_0-(\Gamma-\Gamma_0)$ precisely with no prefactors.
We explain this particularly simple, striking behavior below.
We plot both $\gamma_*^+$, the yield point in the forward direction, and $-\gamma_*^-$, the negative of the yield point in the backward direction, but the values are identical to within the size of the symbols.

In figure~\ref{fig:fig3}b) we plot $\sigma_*^+$, the stress at forward onset (pluses), $-\sigma_*^-$, the negative of the stress at reverse onset (crosses), and their average, $\sigma_*$ (dots).
In c) we do the same for the plateau value of plastic strain, $\epsilon_p^\pm$, at the forward and reverse turning points.
In both b) and c), there is some $\pm$ asymmetry for smaller $\Gamma$, but this effect goes away at larger $\Gamma$.
The $\sigma_*$ curve in b) approaches a $\frac{d\sigma_*}{d\Gamma}\approx -1/2$ behavior at larger $\Gamma$.
Below $\Gamma_0=0.36320$, $\epsilon_p^{+}=\epsilon_p^{-}$ and $\epsilon_p=0$ identically, since there is no hysteresis.
The common value of $\epsilon_p^{\pm}$ has a residual positive sign corresponding to the symmetry breaking from the arbitrary decision to begin shearing the virgin system in the positive direction, and this results in the symmetry breaking between $\sigma_*^+$ and $\sigma_*^-$ since there is no symmetry breaking between $\gamma_*^+$ and $\gamma_*^-$.
In the inset, we plot $\epsilon_p$ vs $\Gamma-\Gamma_0$ on a log-log scale and find a power-law behavior, indicative of a second-order transition, with $\epsilon_p \propto (\Gamma-\Gamma_0)^{1.2\pm 0.1}$.

As $\Gamma$ increases, we find $\frac{d\epsilon_p}{d\Gamma}\approx +1/2$ in almost precise opposition to $\sigma_*$.
We can understand this by neglecting the plasticity accumulated in the creeping regime, and associating $\epsilon$ at forward yield with $\epsilon_p^{-}$ (in terms of Fig.~\ref{fig:fig1}a, we assume the crosses and stars are at the same height).
Then we must have $\frac{d\sigma_*}{d\Gamma}-\frac{d\gamma_*}{d\Gamma}-\frac{d\epsilon_p}{d\Gamma}=0$ which explains the relation between the $\Gamma$ dependence of $\sigma_*$ and $\epsilon_p$ given the $\gamma_*=\Gamma_0-(\Gamma-\Gamma_0)$ behavior.

\begin{figure}[h!]
\includegraphics[width = 1.\columnwidth]{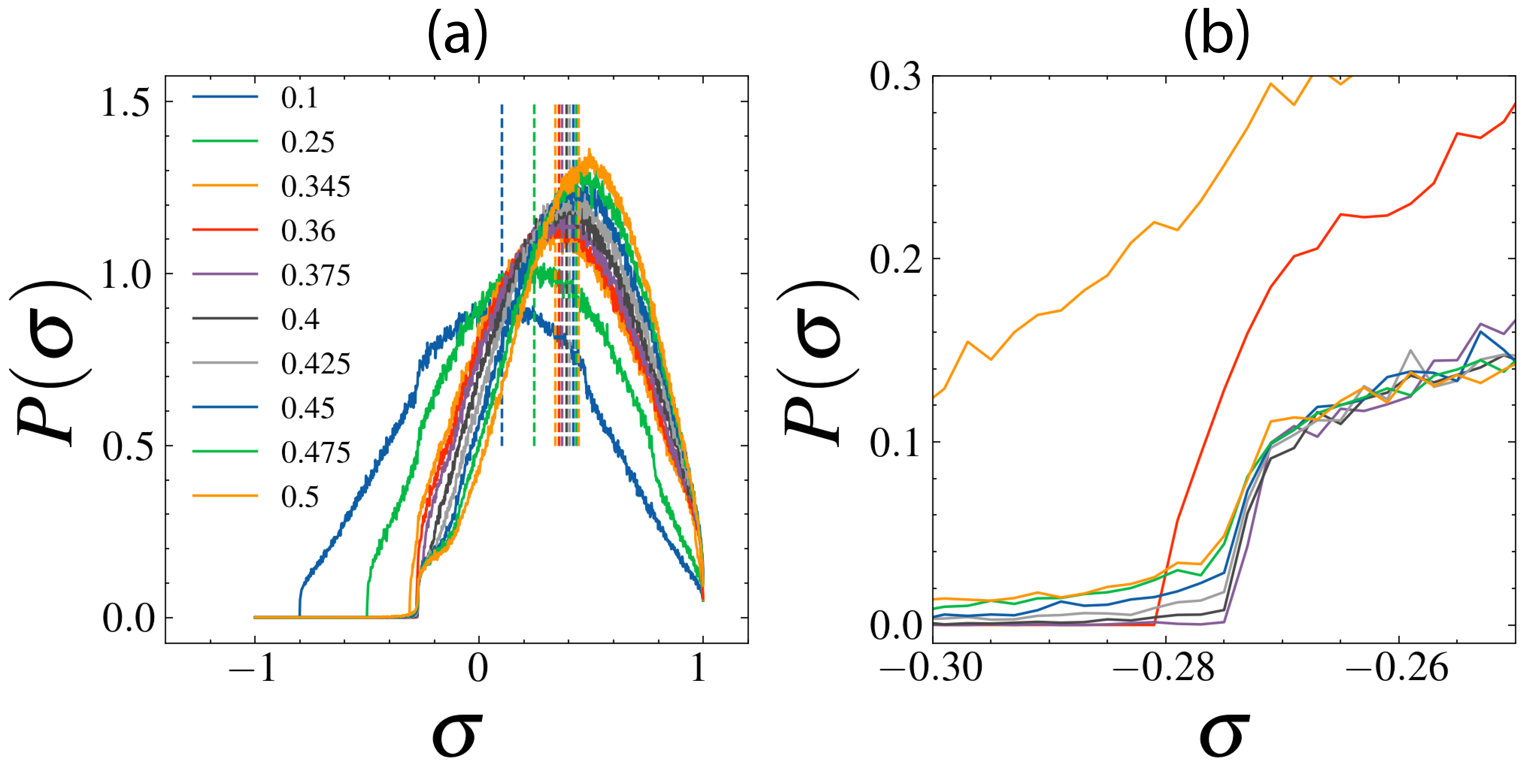}%
\caption{a) The distribution of the stress $\sigma$ at the forward turning point ($\gamma=\Gamma$) for various $\Gamma$.  Vertical dashed lines indicate the average $\sigma$. b) Is simply a zoom on the shoulder region.}
\label{fig:fig4}
\end{figure}
In figure~\ref{fig:fig4}, we show the distribution of local stresses, $P(\sigma)$, at the forward turning point (indicated by the crosses in figure~\ref{fig:fig1}a). 
We show $P(\sigma)$ for several $\Gamma$ values both below and above $\Gamma_0$.
The average $\sigma$ (which corresponds to the macroscopic stress) is indicated by vertical dashed lines.
The reversible plastic systems at $\Gamma>\Gamma_0$ are pronouncedly asymmetric about the average and the most likely stress is noticeably above the average.
If theses systems were subsequently placed into steady forward or reverse shear, the ER systems would likely behave the same in the forward and reverse directions, while the RPR systems would behave differently if one continued shearing indefinitely in the forward direction as opposed to shearing indefinitely in the reverse direction.
As such, the RPR systems are mechanically polarized, while the ER systems are not.~\footnote{The $\pm$ asymmetry in $\sigma_*$ for $\Gamma<\Gamma_0$ is simply due to the fact that the nominal $\gamma=0$ point is not an appropriate stress-free reference state.}
There is also a pronounced shoulder at $\sigma_0=-0.273203$ in all the curves.
The height of the shoulder is at $P_*\approx 0.1$, and this also is essentially independent of $\Gamma$.
As $\Gamma$ increases, the $P(\sigma)$ curves cling to the shoulder further and further up to larger $\sigma$ values before departing.
We now explain the origin of $\sigma_0$, its relation to $\Gamma_0$,  and the nearly prefect $\gamma_*=\Gamma_0-(\Gamma-\Gamma_0)$ behavior.

In our discretized version of Eshelby's problem, a tile, when it transforms, will transition from a stress of $\sigma=1$ to a stress of $\sigma_0=-0.273203$.
This is precisely the $\sigma$ value at which the shoulder appears in the $P(\sigma)$ distribution.
We show the distributions as they exist at the forward turning point, so applying reverse shear, if we temporarily neglect the redistribution of stress from the plastic events which would occur in the creeping regime at $\sigma<\sigma_0$, shifts all stress values uniformly to the left.
For the systems which are below threshold, there is a gap between $\sigma=-1$ and the smallest stress value which is precisely equal to $2\Gamma$.
For instance, the system cycled at $\Gamma=0.1$ (blue curve) has $P=0$ for all $\sigma$ between $-1$ and $-0.8$, and during the cycling, the distribution is simply advected with no change in form, since there is no plasticity.
The system cycled at $\Gamma=0.5$ which is well above $\Gamma_0$ has some residual non-zero $P$ to the left of the shoulder at $\sigma<\sigma_0$,  but these sites  contribute only to the creeping regime below $\gamma_*$.
Primary plasticity initiates when $\gamma=-\gamma_*$, and the shoulder is advected to precisely to $\sigma=-1$.  
The critical amplitude for which the shoulder will just reach $\sigma=-1$ upon reversal would then be $\Gamma_0=(1+\sigma_0)/2=0.363399$ which is in near perfect agreement with our observation of the onset of reversible plasticity between $\Gamma=0.360$ and $\Gamma=0.365$  and the value of $\Gamma_0=0.36320$ inferred from the $\gamma_*$ vs $\Gamma$ curves.

Beyond the values of $\Gamma_0$ and the $\gamma_*$ dependence on $\Gamma$, the $P(\sigma)$ distributions can also tell us more about the shape of the $\epsilon$ vs. $\gamma$ loading curves.
We can associate the fact that when primary plasticity initiates, the plastic strain rate is roughly $0.2$ regardless of $\Gamma$ with the fact that the shoulder height in $P(\sigma)$ is roughly $0.1$ regardless of $\Gamma$.
Each site which yields gives a plastic strain of $2/L^2$, so if we neglect post-yield stress redistribution and assume that the action of loading the system simply depletes the sites at $\sigma=-1$ and re-injects them at $\sigma_0$, then a shoulder height of $0.1$ would precisely give a plastic strain rate of $0.2$ at the onset.
Furthermore, we have checked that integrating $P(\sigma)$ over the creeping regime below the shoulder gives the total creeping strain, i.e. the difference in heights between the crosses and stars, in figure~\ref{fig:fig1}a.

To summarize, we have shown here that the systems farther into the reversible plastic regime are anomalously soft in the sense that despite having lowered energy, they also show plasticity which initiates earlier at lower applied stress and strain.
We have explained this observation, and, in particular, the onset amplitude at which one begins to see reversible plasticity in the steady state, in terms of an accumulation in the stress distribution at the stress level, $\sigma_0$, at which an Eshelby inclusion finds itself immediately after a transformation.
We have also shown that the transition from the elastic to the reversible plastic regime has a mixed first/second order character where the amplitude of the hysteresis increases continuously from zero as the driving amplitude increases, but where the dissipation rate jumps from zero to a finite value at the endpoint of the loading curve associated with a finite residue, $P_*$ in the stress distribution, $P(\sigma)$ at $\sigma_0$.

Our model is essentially the simplest real-space model one could imagine for an amorphous material subjected to athermal quasi-static flow with no adjustable parameters or prescribed disorder.
 We expect that in experiments, in EPMs with a disordered landscape, or in particle-based models, there could be some broadening of the shoulder we see in $P(\sigma)$. 
However, it is likely that the basic mechanism we have identified and the emergence of a characteristic $\sigma_0$ and its connection with the optimal hardness and onset of reversible plasticity at $\Gamma_0$ will be robust, at least for highly disordered initial states.
In fact, in~\cite{fiocco2013oscillatory} and \cite{Leishangthem:2017us} evidence for the anomalous softness was implicit in the stress-strain curves, but no mention was made of it, and no quantification was attempted.
A more precise quantification of the anomalous softness in models, simulations, and experiments should be an important program going forward.

\begin{acknowledgments}
We acknowledge computing support on the discovery computing cluster at the MGHPCC.
\end{acknowledgments}

\bibliography{Bib.bib}

\end{document}


\title{Staggered grid discretization of Eshelby's problem}
\author{A.~Elgailani}
\affiliation{Department of Mechanical and Industrial Engineering,
Northeastern University, Boston, Massachusetts 02115, USA}
\author{D.~Vandembroucq}
\affiliation{PMMH, CNRS UMR 7636, ESPCI Paris, PSL University, Sorbonne  Université, Université  Paris Cité, F-75005 Paris, France}
\author{C.E. ~Maloney}
\affiliation{Department of Mechanical and Industrial Engineering, Northeastern University, Boston, Massachusetts 02115, USA}

\maketitle

The automaton model we use here is almost identical to the one employed in reference~\cite{Khirallah-PRL2021}.
However, we have changed the details of how the elasticity is discretized.
In the present study have used a so-called staggered grid discretization which is a common procedure in seismology~\cite{10.1785/BSSA0660030639,Levander:1988ue,Randall1991MultipoleBA,Virieux:1986tb,10.1785/BSSA0860041091} and has recently started to see more use in the materials science and theoretical solid mechanics communities~\cite{Schneider:2016tn}.
The discretization used here results in precisely the same $K_{11}$ and $K_{22}$ (defined below) as in reference~\cite{Khirallah-PRL2021}, but different $K_{12}$.
The $K_{12}$ used in reference~\cite{Khirallah-PRL2021} had some pathological behavior which is not present in the staggered grid discretization.
Although we do not think these pathologies affected the results in that study or would present any problems in the present study, we nevertheless have opted to use the staggered grid technique here.

We start by defining a displacement field $u_\alpha[I,J]$ on a regular grid with co-ordinates indices $I,J$.
The grid is square with $0\leq I <L$, $0\leq J < L$.
We define the total strain, $\epsilon$, from the displacement field using the staggered grid technique~\cite{Schneider:2016tn} and assuming periodic boundary conditions on $u$ so that $u[I+mL,J+nL]=u[I,J]$ for any integers $m,n$.
This gives:
\begin{equation}
\gamma_{xx}[I,J]=u_x[I+1,J]-u_x[I,J]
\end{equation}
\begin{equation}
\gamma_{xy}[I,J]=u_y[I,J]-u_y[I-1,J]
\end{equation}
\begin{equation}
\gamma_{yx}[I,J]=u_x[I,J]-u_x[I,J-1]
\end{equation}
\begin{equation}
\gamma_{yy}[I,J]=u_y[I,J+1]-u_y[I,J]
\end{equation}
For a physical interpretation of this staggered difference scheme, see reference~\cite{Schneider:2016tn}
Note that not all $\gamma$ fields are derivable from a $u$ field.
The $\gamma$ fields which \emph{are} derived from a $u$ field this way are kinematically compatible within our discretization scheme.
We can define a local elastic strain energy as
\begin{equation}
\phi[I,J]=\frac{\mu}{2}\left((\gamma_1[I,J]-\epsilon_1[I,J])^2+(\gamma_2[I,J]-\epsilon_2[I,J])^2\right)+\frac{K}{2}(\gamma_3[I,J])^2
\end{equation}
where $\mu$ and $K$ are the shear and compression moduli, $\gamma_1$ and $\gamma_2$ are the two shear components and $\gamma_3$ the dilatant component of the total strain:
\begin{equation}
\gamma_1=\gamma_{xx}-\gamma_{yy}
\end{equation}
\begin{equation}
\gamma_2=\gamma_{xy}+\gamma_{yx}
\end{equation}
\begin{equation}
\gamma_3=\gamma_{xx}+\gamma_{yy}
\end{equation}
and $\epsilon_1$ and $\epsilon_2$ are two prescribed plastic strain fields or so-called eigenstrains.
The total elastic strain energy is then simply a sum over the lattice.
\begin{equation}
\phi=\sum_{IJ}\phi[I,J]
\end{equation}
The discretized Eshelby problem is then to find a displacement field and the associated strain field which minimizes $\phi$ subject to the prescribed $\epsilon_1$ and $\epsilon_2$.
Note that posing the minimization problem in terms of displacements automatically ensures the resulting strain field is kinematically compatible.
If the prescribed $\epsilon$ field happens to be compatible, then the solution will be trivial with $\gamma=\epsilon$, but we are typically interested in incompatible $\epsilon$ fields.
Because of the linearity of the problem, we can express the solution for any arbitrary prescribed $\epsilon_1$ and $\epsilon_2$ fields in terms of the solution to a unit delta source, 
\begin{equation}
\epsilon_1[I,J]=\delta_{I,0}\delta_{J,0}
\end{equation}
and
\begin{equation}
\epsilon_2[I,J]=\delta_{I,0}\delta_{J,0}.
\end{equation}
Note that these delta source fields are not kinematically compatible and could not have been derived from any displacement field.
We define the strain fields derived from the displacements resulting from the energy minimization problem with a delta source as $K_{11}, K_{12}, K_{21}, K_{22}$ where: $K_{11}$ is the $\gamma_1$ strain resulting from an imposed $\epsilon_1$ eigenstrain,  $K_{12}$ is the $\gamma_1$ strain resulting from an imposed $\epsilon_2$ eigenstrain, $K_{21}$ is the $\gamma_2$ strain resulting from an imposed $\epsilon_1$ eigenstrain, and $K_{22}$ is the $\gamma_2$ strain resulting from an imposed $\epsilon_2$ eigenstrain.
We specialize here to the incompressible limit where $K/\mu\rightarrow\infty$.
The solution to this energy minimization problem is most easily expressed in Fourier space.
\begin{eqnarray}
\tilde{K}_{11}[p,q]&=&-\frac{4q_{x-}q_{x+}q_{y-}q_{y+}}{\mathcal{D}}=-\frac{4\Delta^2_x\Delta^2_y}{\mathcal{D}}\\
\tilde{K}_{22}[p,q]&=&-\frac{(q_{x-}q_{x+}-q_{y-}q_{y+})^2}{\mathcal{D}}=-\frac{(\Delta^2_x-\Delta^2_y)^2}{\mathcal{D}}\\
\tilde{K}_{12}[p,q]&=&\frac{2 q_{x+} q_{y+} (q_{x-} q_{x+} - q_{y-} q_{y+})}{\mathcal{D}}=\frac{2 q_{x+} q_{y+} (\Delta^2_x-\Delta^2_y)}{\mathcal{D}}\\
\tilde{K}_{21}[p,q]&=&\frac{2 q_{x-} q_{y-} (q_{x-} q_{x+} - q_{y-} q_{y+})}{\mathcal{D}}=\frac{2 q_{x-} q_{y-} (\Delta^2_x-\Delta^2_y)}{\mathcal{D}}
\end{eqnarray}
where
\begin{eqnarray}
\Delta^2_x&=&q_{x+}q_{x-}\\
\Delta^2_y&=&q_{y+}q_{y-}
\end{eqnarray}
are the Fourier transforms of the second ordered centered difference operators in the $x$ and $y$ direction and where
\begin{equation}
\mathcal{D}=(\Delta_x^2+\Delta_y^2)^2
\end{equation}
is the Fourier transform of the the graph Laplacian of the graph Laplacian, 
and where 
\begin{eqnarray}
q_{x+}[p,q]&=&-i (+\exp[+2\pi i p/L]-1)\\
q_{x-}[p,q]&=&-i (-\exp[-2\pi i p/L]+1)\\
q_{y+}[p,q]&=&-i (+\exp[+2\pi i q/L]-1)\\
q_{y-}[p,q]&=&-i (-\exp[-2\pi i q/L]+1)\\
\end{eqnarray}
are the Fourier transforms of the forward and backward difference operators in the $x$ and $y$ directions.
We have used the discrete Fourier transform conventions:
\begin{equation}
\tilde{K}[p,q]=\left(\frac{1}{\sqrt{L}}\right)^2\sum_{I,J}K[I,J]\exp\left[+2\pi (Ip+Jq)/L \right]
\end{equation}
\begin{equation}
K[I,J]=\left(\frac{1}{\sqrt{L}}\right)^2\sum_{p,q}\tilde{K}[p,q]\exp\left[-2\pi (Ip+Jq)/L \right]
\end{equation}
With the solution for the delta sources, we can then write the solution for an arbitrary source field in terms of a convolution:
\begin{equation}
\gamma_{1}[I,J]=\sum_{IJ}K_{11}[M-I,N-J]\epsilon_{1}[M,N] + K_{12}[M-I,N-J]\epsilon_{2}[M,N]
\end{equation}
\begin{equation}
\gamma_{2}[I,J]=\sum_{IJ} K_{21}[M-I,N-J]\epsilon_{1}[M,N] + K_{22}[M-I,N-J]\epsilon_{2}[M,N]
\end{equation}
Then, of course, because of the convolution theorem, we have a decoupled mode-wise relation in for each $p,q$ mode in Fourier space,
\begin{equation}
\tilde{\gamma}_{1}[p,q]=\tilde{K}_{11}[p,q]\tilde{\epsilon}_{1}[p,q] + \tilde{K}_{12}[p,q]\tilde{\epsilon}_{2}[p,q]
\end{equation}
\begin{equation}
\tilde{\gamma}_{2}[p,q]=\tilde{K}_{21}[p,q]\tilde{\epsilon}_{1}[p,q] + \tilde{K}_{22}[p,q]\tilde{\epsilon}_{2}[p,q]
\end{equation}
Our expressions for $K$ were obtained by differentiating the energy with respect to $u_x$ and $u_y$, solving the linear equations for the $u$ fields which give zero energy derivative and then re-inserting those $u$ fields back into the staggered difference scheme to find $\gamma$, and we have not shown those steps here as they are straightforward but tedious.
We note that the present $\tilde{K}_{11}$ and $\tilde{K}_{22}$ are \emph{precisely} the same as in reference~\cite{Khirallah-PRL2021} and as in the continuum Eshelby problem~\cite{picard2004elastic}.
It is only $\tilde{K}_{12}$ and $\tilde{K}_{21}$ which are different in the present staggered-difference scheme. 
The definition of the automaton from a piece-wise quadratic strain-energy function and the initialization procedure are then precisely as in reference~\cite{Khirallah-PRL2021} so we do not elaborate further here. 
\bibliography{Bib.bib}